\documentclass[a4paper,12pt]{article}
\usepackage{amssymb}
\usepackage[dvips]{graphicx}
\usepackage[usenames]{color}
\usepackage{amsmath}
\usepackage{fullpage}
\usepackage{latexsym}

\makeatletter \@addtoreset{equation}{section} \makeatother

\begin{document}

\renewcommand{\[}{\begin{equation}} \renewcommand{\]}{\end{equation}} %
\renewcommand{\>}{\rangle}

\begin{titlepage}

    \thispagestyle{empty}
    \begin{flushright}
        \hfill{CERN-PH-TH/2012-006}\\
    \end{flushright}

    \vspace{15pt}
    \begin{center}
        { \huge{\textbf{Quantum Gravity\\\vspace{8pt} Needs Supersymmetry}}}\vspace{25pt}

        \vspace{55pt}

         {\large{\bf Sergio Ferrara$^{1,2}$ and\ Alessio Marrani$^{1}$}}

        \vspace{15pt}

        {$1$ \it Physics Department,Theory Unit, CERN, \\
        CH 1211, Geneva 23, Switzerland\\
        \texttt{sergio.ferrara@cern.ch}\\
        \texttt{alessio.marrani@cern.ch}}

        \vspace{10pt}

        {$2$ \it INFN - Laboratori Nazionali di Frascati, \\
        Via Enrico Fermi 40,00044 Frascati, Italy}

 \vspace{100pt}

 \begin{abstract}
We report on some old and new results on the quantum aspects of
four-dimensional maximal supergravity, and its hypothetical
ultraviolet finiteness.
\end{abstract}

\vspace{130pt}

        \noindent \textit{Contribution to the Proceedings of the International School of Subnuclear
        Physics,\\\vspace{2pt}49th Course: ``Searching for the Unexpected at LHC and Status of Our Knowledge",\\\vspace{2pt}Erice, Italy, June 24 -- July 3, 2011\\Based on a lecture given by S. Ferrara}
\end{center}

\end{titlepage}
\newpage

\section{From Supersymmetry...}

If supersymmetry is used in \textit{string theory} (see \textit{e.g.} \cite%
{Schwarz,Polchinski}), then it is possible to construct five seemingly
consistent ($M$-theory descendant \cite{Witten}; see \textit{e.g.} \cite%
{Duff-rev} for a list of Refs.) theories in $D=10$ space-time dimensions
(see \textit{e.g.} \cite{GSW}). However, due to the large possible choice of
compactification manifolds
\begin{equation}
\mathcal{M}_{10}\longrightarrow \mathcal{M}_{4}\times X_{6},
\end{equation}%
a multitude of $D=4$ theories arise (it can be classified \textit{e.g.} by a
statistical approach to the vacua \cite{Douglas}), and it is yet a mystery
how string theory could select the physical vacuum. Further models involving
physical degrees of freedom living in the bulk and on branes give very
interesting possibilities (see \textit{e.g.} \cite%
{RS,Antoniadis,Dvali,Dimopoulos}).

On the other hand, if supersymmetry is used in \textit{field theory}, then
one encounters the opposite situation. For \textit{rigid} theories (\textit{%
i.e.}, neglecting gravity), one can build nice but unrealistic models,
explore AdS/CFT duality, and study strongly coupled gauge theories. However,
when gravity is included, \textit{at least} in the cases with $\mathcal{N}%
\leqslant 16$ supercharges, it seems that one fails to unify supersymmetry,
gravity and quantum theory.

In spite of these drawbacks, as we will shortly report in the next Sections,
there are some indications that for some $D=4$ special theories of
supergravity, namely $\mathcal{N}=8$ and possibly $\mathcal{N}=5$, $6$,
\textquotedblleft miracles\textquotedblright\ occur due to unexpected
cancellations in multi-loop perturbative calculations \cite%
{4-loop,BHS,BHN,K-1,Vanhove,Dixon,K-2,Freedman,Bern-last}.

Arguments based on $M$-theory compactifications \cite%
{Green:1997di,Green:2010sp,Bjornsson:2010wm} and $E_{7}$ invariance \cite%
{Freedman} prevent counterterms only up to a finite loop order. On the other
hand, analysis based on possible divergences in the supefield light-cone
formulation may indicate the absence of counterterms at all loops \cite%
{K-1,K-2}.

In October 1976, S.F. was invited by John Schwarz to visit Caltech and to
give a seminar on the construction of the first supersymmetric theory of
gravity in $D=4$ (named $\mathcal{N}=1$ \textit{supergravity} \cite{N=1,DZ}%
); at that time, the first extended ($\mathcal{N}=2)$ supergravity had just
been completed \cite{N=2}. Murray Gell-Mann, during an enlightening
conversation in his office, remarked that if $D=4$ higher-$\mathcal{N}$
supergravities existed, then $\mathcal{N}=8$ would be the supergravity
theory with maximal supersymmetry \cite{GMS}.

Today, after 35 years, we are still struggling with $\mathcal{N}=8$ \textit{%
maximal} supergravity, its connections to superstring and $M$- theory, its
hypothetical perturbative finiteness and its non-perturbative completion(s).

\section{...to Maximal Supergravity...}

As anticipated by Gell-Mann, $\mathcal{N}=8$ supergravity \cite%
{Cremmer:1979up,dWN} is the theory with the largest possible amount of
supersymmetry for particles with spin $s\leqslant 2$ in $D=4$ (namely, no
higher spin fields in the massless spectrum). Indeed, in supersymmetric
gravity theories with $\mathcal{N}$-extended supersymmetry, the massless
particle content is given by
\begin{equation}
\binom{\mathcal{N}}{k}\equiv \frac{\mathcal{N}!}{k!\left( \mathcal{N}%
-k\right) !}\text{ particles~of~\textit{helicity}~}\lambda =2-\frac{k}{2},
\end{equation}
where $k_{\max }=\mathcal{N}$, and $\mathcal{N}\leqslant 8$ if $\left|
\lambda \right| \leqslant 2$ is requested.

One possible approach to maximal supergravity is to consider it as it comes
from $M$-theory restricted to the massless sector. The problem is that this
theory, even if preserving maximal $\mathcal{N}=8$ supersymmetry
(corresponding to $32=8\times 4$ supersymmetries), is \textit{not} uniquely
defined, because of the multiple choice of internal compactification
manifolds and corresponding duality relations:
\begin{equation}
\begin{array}{ll}
\begin{array}{l}
\mathbf{I}.~M_{11}\longrightarrow M_{4}\times T_{7} \\
~%
\end{array}
&
\begin{array}{l}
\text{(}GL^{+}(7,\mathbb{R})\text{ and }SO\left( 7\right) ~\text{manifest);}
\\
~%
\end{array}
\\
\begin{array}{l}
\mathbf{II}.~M_{11}\longrightarrow AdS_{4}\times S^{7} \\
~%
\end{array}
&
\begin{array}{l}
\text{(}SO\left( 8\right) ~\text{manifest,~\textit{gauged});} \\
~%
\end{array}
\\
\mathbf{III}.~M_{11}\longrightarrow M_{4}\times T_{7,\mathcal{R}} & \text{(}%
SL(8,\mathbb{R})\text{ and }SO\left( 8\right) ~\text{manifest),}%
\end{array}
\label{various-N=8,d=4}
\end{equation}
where $T_{7}$ is the $7$-torus and $S^{7}$ is the $7$-sphere. $T_{7,\mathcal{%
R}}$ denotes the case in which, according to Cremmer and Julia \cite%
{Cremmer:1979up}, the dualization of $21$ vectors and $7$ two-forms makes $%
SL(8,\mathbb{R})$ (in which $GL^{+}(7,\mathbb{R})$ is maximally embedded)
manifest as maximal non-compact symmetry of the Lagrangian. Note that in
case $\mathbf{III}$ one can further make $E_{7\left( 7\right) }$ (and its
maximal compact subgroup $SU\left( 8\right) $) manifest \textit{on-shell},
by exploiting a Cayley transformation supplemented by a rotation through $%
SO\left( 8\right) $ gamma matrices on the vector\ $2$-form field strengths
\cite{Cremmer:1979up,HW}.

The fundamental massless fields (and the related number $\sharp $ of degrees
of freedom) of $M$-theory in $d=11$ flat space-time dimensions are \cite{CSJ}
\begin{equation}
\begin{array}{lll}
\begin{array}{l}
g_{\mu \nu }~\text{(\textit{graviton})}: \\
~%
\end{array}
&
\begin{array}{l}
\sharp =\frac{\left( d-1\right) \left( d-2\right) }{2}-1, \\
~%
\end{array}
&
\begin{array}{l}
\text{in~}d=11:\sharp =44; \\
~%
\end{array}
\\
\begin{array}{l}
\Psi _{\mu \alpha }~\text{(\textit{gravitino})}: \\
~%
\end{array}
&
\begin{array}{l}
\sharp =(d-3)2^{(d-3)/2}, \\
~%
\end{array}
&
\begin{array}{l}
\text{in~}d=11:\sharp =128; \\
~%
\end{array}
\\
A_{\mu \nu \rho }~\text{(\textit{three-form})}: & \sharp =\frac{\left(
d-2\right) \left( d-3\right) \left( d-4\right) }{3!}, & \text{in~}%
d=11:\sharp =84.%
\end{array}%
\end{equation}%
Because a $\left( p+1\right) $-form (\textquotedblleft
Maxwell-like\textquotedblright\ gauge field) $A_{p+1}$ couples to $p$%
-dimensional extended objects, and its \textquotedblleft
magnetic\textquotedblright\ dual $B_{d-p-3}$ couples to $\left( d-p-4\right)
$-dimensional extended objects, it follows that the fundamental (massive)
objects acting as sources of the theory are $M2$- and $M5$-branes.

In the formulation $\mathbf{III}$ of (\ref{various-N=8,d=4}) \cite%
{Cremmer:1979up}, the gravitinos $\psi _{I}$ and the gauginos $\chi _{IJK}$
respectively have the following group theoretical assignment\footnote{%
As evident from (\ref{fermions-III}), we use a different convention with
respect to \cite{Slansky} (see \textit{e.g.} Table 36 therein). Indeed, we
denote as $\mathbf{8}_{v}$ of $SO\left( 8\right) $ the irrep. which
decomposes into $\mathbf{7}+\mathbf{1}$ of $SO\left( 7\right) $, whereas the
two spinorial irreps. $\mathbf{8}_{s}$ and $\mathbf{8}_{c}$ both decompose
into $\mathbf{8}$ of $SO\left( 7\right) $. The same change of notation holds
for $\mathbf{35}$ and $\mathbf{56}$ irreps..} ($I$ in $\mathbf{8}$ of $%
SU\left( 8\right) $):
\begin{equation}
\text{theory }\mathbf{III~}\text{\cite{Cremmer:1979up}}:\left\{
\begin{array}{l}
\psi _{I}:\underset{\mathbf{8}}{~SO\left( 7\right) }\subset \underset{%
\mathbf{8}_{s}}{SO\left( 8\right) }\subset \underset{\mathbf{8}}{SU\left(
8\right) }; \\
\\
\chi _{IJK}:\underset{\mathbf{8}+\mathbf{48}}{~SO\left( 7\right) }\subset
\underset{\mathbf{56}_{s}}{SO\left( 8\right) }\subset \underset{\mathbf{56}}{%
SU\left( 8\right) }.%
\end{array}
\right.  \label{fermions-III}
\end{equation}
On the other hand, the $70$ scalar fields arrange as
\begin{equation}
\text{theory }\mathbf{III~}\text{\cite{Cremmer:1979up}}:\underset{\left(
\sharp =70\right) }{s=0\text{~\textit{dofs}}}:~\underset{\mathbf{1}+\mathbf{7%
}+\mathbf{21}+\mathbf{35}}{SO\left( 7\right) }\subset \underset{\mathbf{35}%
_{v}+\mathbf{35}_{c}}{SO\left( 8\right) }\subset \underset{\mathbf{70}}{%
SU\left( 8\right) },  \label{scalars-III}
\end{equation}
where $\mathbf{70}$ is the rank-$4$ completely antisymmetric irrep. of $%
SU\left( 8\right) $, the maximal compact subgroup of the $U$-duality group $%
E_{7\left( 7\right) }$ (also called $\mathcal{R}$-symmetry). It follows that
scalars parameterize a non-compact coset manifold $\frac{G}{SU\left(
8\right) }$. Indeed, the $SU\left( 8\right) $ under which both the scalar
fields and the fermion fields transform is the \textit{``local''} $SU\left(
8\right) $, namely the stabilizer of the scalar manifold. On the other hand,
also a \textit{``global''} $SU\left( 8\right) $ ($\mathcal{R}$-symmetry
group) exists, under which the vector $2$-form self-dual/anti-self-dual
field strengths transform. Roughly speaking, the physically relevant group $%
SU\left( 8\right) $ is the diagonal one in the product $SU_{\text{local}%
}\left( 8\right) \times SU_{\text{global}}\left( 8\right) $ (see also
discussion below).

Remarkably, there exists an \textit{unique} simple, non-compact Lie group
with real dimension $70+63=133$ and which embeds $SU\left( 8\right) $ as its
\textit{maximal} compact subgroup: this is the real, non-compact split form $%
E_{7\left( 7\right) }$ of the exceptional Lie group $E_{7}$, thus giving
rise to the symmetric, rank-$7$ coset space
\begin{equation}
\frac{E_{7\left( 7\right) }}{SU\left( 8\right) /\mathbb{Z}_{2}},
\end{equation}
which is the scalar manifold of $\mathcal{N}=8$, $D=4$ supergravity ($%
\mathbb{Z}_{2}$ is the kernel of the $SU\left( 8\right) $-representations of
even rank; in general, spinors transform according to the double cover of
the stabilizer of the scalar manifold; see \textit{e.g.} \cite%
{Yokota,AFZ-rev}).

$E_{7\left( 7\right) }$ acts as electric-magnetic duality symmetry group
\cite{GZ}, and its maximal compact subgroup $SU\left( 8\right) $ has a
chiral action on fermionic as well as on (the vector part of the) bosonic
fields. While the chiral action of $SU\left( 8\right) $ on fermions directly
follows from the chirality (complex nature) of the relevant irreps. of $%
SU\left( 8\right) $ (as given by Eq. (\ref{fermions-III})), the chiral
action on vectors is a crucial consequence of the electric-magnetic duality
in $D=4$ space-time dimensions. Indeed, this latter allows for \textit{%
``self-dual / anti-self-dual''} complex combinations of the field strengths,
which can then fit into complex irreps. of the stabilizer $H$ of the coset
scalar manifold $G/H$ itself. For the case of maximal $\mathcal{N}=8$
supergravity, the relevant chiral complex irrep. of $H=SU\left( 8\right) $
is the rank-$2$ antisymmetric $\mathbf{28}$.

Note that if one restricts to the $SL\left( 8,\mathbb{R}\right) $-covariant
sector, the chirality of the action of electric-magnetic duality is spoiled,
because the maximal compact subgroup of $SL\left( 8,\mathbb{R}\right) $,
namely $SO\left( 8\right) $, has not chiral irreps.

Composite (sigma model $G/H$) anomalies can arise in theories in which $G$
has a maximal compact subgroup with a \textit{chiral} action on bosons
and/or fermions (see \textit{e.g.} \cite{DFG,Marcus,BHN}). Surprising
cancellations among the various contributions to the composite anomaly can
occur as well. An example is provided by $\mathcal{N}=8$, $d=4$ supergravity
itself, in which standard anomaly formul\ae\ yield the remarkable result
\cite{Marcus,BHN}
\begin{equation}
3Tr_{\mathbf{8}}X^{3}-2Tr_{\mathbf{28}}X^{3}+Tr_{\mathbf{56}}X^{3}=\left(
3-8+5\right) Tr_{\mathbf{8}}X^{3}=0,  \label{(caN=8)=0}
\end{equation}
where $X$ is any generator of the Lie algebra $\mathfrak{su}(8)$ of the
\textit{rigid} (\textit{i.e.} \textit{global)} $SU(8)$ group ($\mathcal{R}$%
-symmetry). In light of the previous considerations, the first and third
contributions to (\ref{(caN=8)=0}) are due to fermions: the $8$ gravitinos $%
\psi _{A}$ and the $56$ spin-$\frac{1}{2}$ fermions $\chi _{ABC}$,
respectively, whereas the second contribution is due to the $28$ chiral
vectors. Note that, for the very same reason, the \textit{local} $SU(8)$
(stabilizer of the non linear sigma-model of scalar fields), under which
only fermions do transform\footnote{%
Also scalar fields transform under \textit{local} $SU\left( 8\right) $, but
they do not contribute to the composite anomaly, because they sit in the
\textit{self-real} (and thus \textit{non-chiral}) rank-$4$ antisymmetric
irrep. $\mathbf{70}$ of $SU\left( 8\right) $.}, would be \textit{anomalous}
\cite{DFG}.\texttt{\ }In an analogous way, in \cite{Marcus} it was
discovered that $\mathcal{N}=6$ and $\mathcal{N}=5$ ``pure'' supergravities
are \textit{composite} \textit{anomaly-free}, whereas $\mathcal{N}\leqslant
4 $ theories are not.

\section{...and its UV Perturbative Finiteness?}

At homotopical level, the following holds:
\begin{equation}
E_{7\left( 7\right) }\cong \left( SU\left( 8\right) /\mathbb{Z}_{2}\right)
\times \mathbb{R}^{70},
\end{equation}
implying that the two group manifolds have the same De Rham cohomology. This
is a key result, recently used in \cite{BHN} to show that the aforementioned
absence of $SU(8)$ current anomalies yield to the absence of anomalies for
the non-linearly realized $E_{7\left( 7\right) }$ symmetry, thus implying
that the $E_{7\left( 7\right) }$ continuous symmetry of classical $\mathcal{N%
}=8$, $d=4$ supergravity is preserved at all orders in perturbation theory
(see \textit{e.g.} \cite{4-loop,K-1,BFK,Vanhove,Dixon,K-2,Freedman}). This
implies the perturbative finiteness of supergravity at least up to seven
loops; Bern, Dixon \textit{et al.} explicitly checked the finiteness up to
four loops included \cite{4-loop} (computations at five loops, which might
be conclusive, are currently in progress; for a recent review, see \textit{%
e.g.} \cite{Bern-last}).

A puzzling aspect of these arguments is that string theory certainly
violates continuous $E_{7\left( 7\right) }$ symmetry at the perturbative
level, as it can be easily realized by considering the dilaton dependence of
loop amplitudes (see \textit{e.g.} \cite{Freedman}). However, this is not
the case for $\mathcal{N}=8$ supergravity. From this perspective, two
(perturbatively finite) theories of quantum gravity would exist, with $32$
local supersymmetries; expectedly, they would differ at least in their
non-perturbative sectors, probed \textit{e.g.} by black hole solutions.
String theorists \cite{GOS,Ark,Banks} claim that $\mathcal{N}=8$, $d=4$
supergravity theory is probably not consistent at the non-perturbative
level. From a purely $d=4$ point of view, their arguments could be overcome
by excluding from the spectrum, as suggested in \cite{BFK}, black hole
states which turn out to be singular or ill defined if interpreted as purely
four-dimensional gravitational objects. Inclusion of such singular states
(such as $\frac{1}{4}$-BPS and $\frac{1}{2}$-BPS black holes) would then
open up extra dimensions, with the meaning that a non-perturbative
completion of $\mathcal{N}=8$ supergravity would lead to string theory \cite%
{GOS}. Extremal black holes with a consistent $d=4$ interpretation may be
defined as having a Bertotti-Robinson \cite{bertotti} $AdS_{2}\times S^{2}$
near-horizon geometry, with a non-vanishing area of the event horizon. In $%
\mathcal{N}=8$ supergravity, these black holes are\footnote{%
We also remark that these are the only black holes for which the \textit{%
Freudenthal duality} \cite{Duff-Freud-1,FD-1} is well defined.} $\frac{1}{8}$%
-BPS or non-BPS (for a recent review and a list of Refs., see \textit{e.g.}
\cite{ICL-1}). The existence of such states would in any case break the $%
E_{7\left( 7\right) }\left( \mathbb{R}\right) $ continuous symmetry, because
of Dirac-Schwinger-Zwanziger dyonic charge quantization conditions.

The breaking of $E_{7\left( 7\right) }\left( \mathbb{R}\right) $ into an
arithmetic subgroup $E_{7\left( 7\right) }\left( \mathbb{Z}\right) $ \cite%
{Hull:1994ys} would then manifest only in exponentially suppressed
contributions to perturbative amplitudes (see \textit{e.g.} the discussion
in \cite{BHN}, and Refs. therein), in a similar way to instanton effects in
non-Abelian gauge theories. Indeed, as for the $SL(2,\mathbb{R})$ symmetry
of Type IIB superstrings in $D=10$ \cite{Green:1997di}, the continuous
symmetry is believed to be broken down to discrete $E_{7\left( 7\right)
}\left( \mathbb{Z}\right) $ by non-perturbative effects like instantons \cite%
{Bianchi:1998nk}. The breaking of $E_{7\left( 7\right) }\left( \mathbb{R}%
\right) $ to $E_{7\left( 7\right) }\left( \mathbb{Z}\right) $ is
instrumental to setting a uniform mass-gap, given by Planck Mass $M_{Pl}=%
\sqrt{\hbar c/G_{N}}=1.22\times 10^{19}GeV/c^{2}$, everywhere inside the
moduli space $E_{7(7)}/SU(8)$, for all regular BPS states with $\mathcal{I}%
_{4}\neq 0$ where \cite{Kallosh:1996uy,Cvetic}
\begin{equation}
\mathcal{I}_{4}=\mathbb{K}_{MNPQ}Q^{M}Q^{N}Q^{P}Q^{Q}  \label{I4}
\end{equation}%
is the \textit{quartic} Cartan invariant of $E_{7(7)}$ \cite{Cremmer:1979up}
and $Q^{M}$ is a $56$-dimensional vector of `bare' quantized electric and
magnetic charges with respect to the 28 $U(1)$ gauge groups.

In \cite{BFK}, after taking into account charge quantization and assuming
that the perturbative theory be UV finite to all orders, the plausibility of
a \textit{non-perturbative} completion of genuinely $D=4$, $\mathcal{N}=8$
supergravity, only including regular black hole states with $\mathcal{I}%
_{4}\neq 0$ and excluding all singular states with $\mathcal{I}_{4}=0$, was
proposed. This proposal has some analogy with $\mathcal{N}=4$ super
Yang-Mills (SYM) theory in $D=4$, decoupled from gravity and other stringy
interactions. Even after including non-perturbative effects, $\mathcal{N}%
=4=4 $ SYM in $D=4$ should not be thought of as a compactification of Type I
or Heterotic strings, that contain the same massless states but differ by
the massive completion, but rather in terms of the AdS/CFT correspondence
\cite{Maldacena:1997re}. Analogously, pure $D=4$ $\mathcal{N}=8$
supergravity, including \textit{regular} non-perturbative states, may be
disconnected from toroidal compactifications of Type II superstrings, that
unavoidably give rise to 1/2 BPS states with $\mathcal{I}_{4}=0$. The fact
that all known $\mathcal{N}=8$ supergravity perturbative amplitudes could be
expressed in terms of $\mathcal{N}=4$ SYM amplitudes in the superconformal
phase, where the latter enjoys $32$ supersymmetries (16 of Poincar\'{e} type
plus 16 superconformal), might be more than an analogy in this respect.

The conjectured UV finiteness of $\mathcal{N}=8$ supergravity, associated
with continuous $E_{7(7)}$ symmetry, has been questioned by Green, Ooguri,
Schwarz in \cite{Green:2007zzb}, where non-decoupling of BPS states from
four-dimensional $\mathcal{N}=8$ supergravity was discussed. The main
conclusion of \cite{Green:2007zzb} was that the $\mathcal{N}=8$ supergravity
limit of string theory does not exist in four dimensions, irrespective of
whether or not the perturbative approximation is free of UV divergences.
String theory adds to the 256 massless states of four-dimensional $\mathcal{N%
}=8$ supergravity an infinite tower of states, such as Kaluza-Klein momenta
and monopoles, wound strings and wrapped branes.

As mentioned, classical solutions of the $\mathcal{N}=8$ version of
non-linear Einstein equations including stable, zero temperature, extremal,
BPS and non-BPS charged black holes \cite{Ferrara:1997tw}, should play a
crucial role in defining the quantum features of the theory. For appropriate
choices of the charges, these black holes can be viewed as smooth solitons
interpolating between flat Minkowski space-time at infinity and
Bertotti-Robinson $AdS_{2}\times S^{2}$ geometry \cite{bertotti} near the
horizon. The asymptotic values of the scalar fields are largely arbitrary
and determine the ADM mass $M$ \cite{ADM} for given charges $Q^{M}$. Thanks
to the \textit{attractor mechanism }\cite{Ferrara:1995ih}, their
near-horizon values are determined in terms of the charges. The entropy of a
black hole in $\mathcal{N}=8$ supergravity is related to the horizon area by
the Bekenstein-Hawking formula
\begin{equation}
S_{BH}={\frac{1}{4}}A_{H}=\pi \sqrt{|\mathcal{I}_{4}|},  \label{quartic}
\end{equation}
and $\mathcal{N}=8$ attractors were studied in \cite{Ferrara:2006em}.

\ Within this framework, the ADM mass of an $\mathcal{N}=8$ extremal black
hole depends on its charges and on the asymptotic values of the scalar
fields, both transforming under $E_{7(7)}$ symmetry:
\begin{equation}
M_{BH}=M_{ADM}(Q,\phi ).
\end{equation}
A manifestly $E_{7(7)}$ covariant expression for the mass is related to the
maximal eigenvalue of the central charge matrix \cite{Hull:1994ys},\cite%
{Ceresole:1995jg}
\begin{equation}
M_{ADM}^{2}(Q,\phi )\geq \text{Max}_{i}\{|Z_{i}(Q,\phi )|^{2}\},
\label{BPS-bound}
\end{equation}
where $Z_{i}(Q,\phi ),i=1,2,3,4$, are the four (skew) eigenvalues \cite%
{BMZ-Hua-Th} of the `dressed' central charge matrix $Z_{AB}(Q,\phi )$.
Indeed, the positive hermitian matrix $\mathcal{H}_{A}{}^{B}=Z_{AC}\bar{Z}%
^{BC}$ has four \textit{real positive eigenvalues} $|Z_{i}(Q,\phi
)|^{2}=\lambda _{i}$ which, without any loss of generality, can be put in
decreasing order $\lambda _{1}\geq \lambda _{2}\geq \lambda _{3}\geq \lambda
_{4}$. The BPS condition requires that the ADM mass be exactly equal to the
largest eigenvalue of the central charge matrix, i.e. it saturates the bound
(\ref{BPS-bound}). On the other hand, non-BPS extremal black holes have a
mass which is strictly larger than the largest eigenvalue of the central
charge matrix; in such a case $\mathcal{I}_{4}<0$, and the non-BPS extremal
black hole geometry is regular and its mass is never zero : $%
S_{BH,non-BPS}=\pi \sqrt{-\mathcal{I}_{4}}$. Thus, it follows that for
regular black holes ($\mathcal{I}_{4}\neq 0$) the ADM mass is bounded from
below as a function in the moduli space; no ``massless black holes'' can
exist in this case, contrarily to the $\mathcal{N}=2$ \cite{Strom,Banks}
(and $\mathcal{N}=4$ \cite{BFK}) cases.

In the SU(8) covariant `dressed' central charge basis, the quartic invariant
(\ref{I4}) in the area/entropy formula (\ref{quartic}) reads
\begin{equation}
\mathcal{I}_{4}(Q,\phi )=Tr[(Z\bar{Z})^{2}]-{\frac{1}{4}}[Tr(Z\bar{Z})]^{2}+8%
\text{Re}Pf(Z),
\end{equation}
where $Pf(Z)$ denotes the \textit{Pfaffian }of $Z_{AB}$; it should be
remarked that each $SU(8)$ invariant term in this expression depends on the
moduli, but nevertheless the total expression is moduli independent due to $%
E_{7(7)}$ symmetry.

As a final comment, it should be stressed that if the perturbative theory
has some - yet unknown - UV divergences \cite{Green:2010sp, Bjornsson:2010wm}%
, the analysis of the massless black hole solutions and the proposal put
forward in \cite{BFK} may require modifications. However, if there are no UV
divergences in perturbation theory, corrections to the analysis of the
space-time properties of these states are not expected. In particular, the
states in \cite{Green:2007zzb} have been shown to be singular in \cite{BFK}.
There are two reasonable options:\textit{\ i)} these states may be
consistently excluded from the four-dimensional theory and therefore do not
affect UV properties of $\mathcal{N}=8$, $D=4$ supergravity; \textit{ii)}
they can be proven to be required in $D=4$ and affect the perturbative
theory.

While some evidence for the plausibility of the first option was given in
\cite{BFK}, a conclusive answer has not been given yet.

\section{Conclusive Remarks}

If $\mathcal{N}=8$, $D=4$ supergravity turns out to be UV perturbatively
finite, according to Bern \textit{et al. }\cite{4-loop,Bern-last}, it is not
only due to maximal supersymmetry and to perturbatively unbroken $E_{7(7)}$
symmetry, but also to other reasons.

One of these is the \textit{\textquotedblleft double copy"} structure \cite%
{Bern-last-2}, which implies a relation, not only kinematical but also
dynamical, between the square of the $\mathcal{N}=4$ super Yang-Mills
amplitudes and $\mathcal{N}=8$ amplitudes. At loop level, the
\textquotedblleft double copy" properties of amplitudes have been extended
to supergravity theories with $\mathcal{N}\geqslant 4$; in this case, one
copy is given by $\mathcal{N}=4$ Yang-Mills and the other copy is an $%
\mathcal{N}=0,1,2$ Yang-Mills gauge theory, thus giving rise to $\mathcal{N}%
=4,5,6$ supergravity theories. From the analysis of divergences, one is led
to conclude that $\mathcal{N}=6$ and $\mathcal{N}=5$ supergravity may be UV
finite (if $\mathcal{N}=8$ is), while $\mathcal{N}=4$ probably is not. It is
worth remarking that these results are in agreement with the
\textquotedblleft composite anomaly" arguments \cite{Marcus,BHN} for which $%
\mathcal{N}=5,6$ do not exhibit duality anomalies, while $\mathcal{N}=4$
does.

Another interesting aspect \cite{Bern-last} which should be implied by UV
finiteness of $N=8,6,5$ supergravity in $D=4$ dimensions is that their
\textit{gauged} versions should be possibly UV finite, as well. Roughly
speaking, this is related to the fact that \textit{gauging} may be regarded
as a spontaneous soft breaking of an unbroken gauge symmetry, and UV
properties should not be affected by such a spontaneous breaking, as it
happens in the Standard Model of electro-weak interactions.

We have already commented on the difficulties and subtleties related to the
question of whether a point-like non-perturbative completion of $\mathcal{N}%
=8$ supergravity exists. Single-centered BPS black hole states preserving a
large fraction of supersymmetry ($1/2$ or $1/4$) are singular in Einstein
theory of gravity, and thus physically unacceptable. However, such states
may be \textit{\textquotedblleft confined"}, \textit{i.e.} they may only
exist as building blocks of multi-centered black hole configurations; the
viability and physical meaning of such a phenomenon are currently unexplored
issues. On the other hand, from a superstring theory perspective such
singular states are just an indication that the fundamental theory is
ill-defined, and that extra dimensions \textit{and/or} a \textit{%
\textquotedblleft non-local"} structure with minimal length open up in the
non-perturbative quantum regime.

\section*{Acknowledgments}

Enlightening discussions with Zvi Bern and Renata Kallosh are gratefully
acknowledged.

The work of S.F. is supported by the ERC Advanced Grant no. 226455, \textit{%
\textquotedblleft Supersymmetry, Quantum Gravity and Gauge
Fields\textquotedblright } (\textit{SUPERFIELDS}).

\end{document}